\documentclass[a4paper,11pt]{article}
\usepackage{pos}

\title{Flavour Physics Phenomenology with SuperIso}
\ShortTitle{SuperIso}

\author*[a]{S. Neshatpour}
\author[b,c]{F. Mahmoudi}

\affiliation[a]{INFN-Sezione di Napoli,
Via Cintia, 80126 Napoli, Italia}

\affiliation[b]{Universit\'e de Lyon, Universit\'e Claude Bernard Lyon 1, CNRS/IN2P3,\\
Institut de Physique des 2 Infinis de Lyon, UMR 5822, F-69622, Villeurbanne, France}

\affiliation[c]{Theoretical Physics Department, CERN,
CH-1211 Geneva 23, Switzerland}

\emailAdd{neshatpour@na.infn.it}
\emailAdd{nazila@cern.ch}

\abstract{%
We present an overview of \texttt{SuperIso v4.1} which is a public program for the calculation of flavour physics observables.
We give examples of using SuperIso to constrain new physics scenarios with kaon physics and present the implications of rare $B$-decays. 
}

\FullConference{%
  Computational Tools for High Energy Physics and Cosmology (CompTools2021)\\
  22-26 November 2021\\
  Institut de Physique des 2 Infinis (IP2I), Lyon, France
}

\begin{document}
\maketitle

\section{Introduction}
The Standard Model of particle physics has been very successful   in explaining the
phenomena within the energies of the Large Hadron Collider. However, there are theoretical and observational indications that there is physics beyond the Standard Model (BSM). 
In addition to direct searches for new physics (NP) signals at particle accelerators, indirect searches play an important and complementary role in investigating BSM physics and can in principle be sensitive to much shorter scales than direct searches probing energies not directly reachable. 
Flavour observables and precision measurements of rare decays are among the most powerful indirect probes of new physics scenarios.

\texttt{SuperIso} is a public C program~\cite{Mahmoudi:2007vz,Mahmoudi:2008tp,Mahmoudi:2009zz,Neshatpour:2021nbn}  with the purpose of studying new physics by analysing indirect searches mainly via flavour observables.
The \texttt{SuperIso} program can be downloaded at \href{http://superiso.in2p3.fr}{http://superiso.in2p3.fr}.  
Details on the installation and the description of the various subroutines can be found in Ref.~\cite{Neshatpour:2021nbn} or in the manual which can be downloaded from the \texttt{SuperIso} website \href{http://superiso.in2p3.fr/superiso4.1.pdf}{http://superiso.in2p3.fr/superiso4.1.pdf}.

In what follows we first list the observables included in \texttt{SuperIso}, in section~\ref{sec:kaon} we briefly review the recently added kaon observables, in section~\ref{sec:Bdecays} we show the implications of rare $B$-decays and we conclude in section~\ref{sec:conclusion}.

\section{Observables}\label{sec:obs}
\texttt{SuperIso}'s main focus is on indirect searches via flavour physics but it also includes other indirect  observables such as electroweak precision tests via oblique parameters $S, T, U$, and $\rho, \Gamma_z$ as well as the muon anomalous moment $(g_\mu -2)$. 
Furthermore, direct search limits from LEP and Tevatron can also be imposed. 
Moreover, dark matter relic
density and direct and indirect detection rates can be included via 
a separate code, \texttt{SuperIso Relic}~\cite{Arbey:2009gu,Arbey:2011zz,Arbey:2018msw}, which is an extension of \texttt{SuperIso}.

The flavour physics observables included in \texttt{SuperIso} can be categorized into two classes of flavour changing charged current (FCCC) and flavour changing neutral current (FCNC) processes.
The former includes tree-level decays involving neutrinos in the final state and consists of the branching ratios of $B\to \ell \nu$, $B\to D^{*}\ell \nu$, $D_s \to \ell \nu$, $D\to \mu\nu$, and $K \to \mu \nu$.
\texttt{SuperIso} includes most of the measured FCNC transitions where there are a plethora of observables mostly in $B\text{-}$physics and also in kaon physics making it possible to probe different types of NP contributions. 
The FCNC transitions in \texttt{SuperIso} consist of:

\vspace{0.3cm}
\noindent{\bf Radiative decays:}
\vspace{-0.1cm}
\begin{itemize}
 \item Branching Ratio (BR) of the inclusive decays $B\to X_{s,d} \gamma$
 \item BR and isospin asymmetry of $B\to K^* \gamma$
\end{itemize}

\vspace{0.3cm}
\noindent{\bf Leptonic decays:}
\vspace{-0.1cm}
\begin{itemize}
 \item BR of $B_{s,d}\to \ell^+ \ell^-$
 \item BR of $K_{L,S} \to \ell^+ \ell^-$
\end{itemize}

\vspace{0.3cm}
\noindent{\bf Semi-leptonic decays:}
\vspace{-0.1cm}
\begin{itemize}
 \item $B\to X_s \ell^+ \ell^-$: BR and angular observables ($A_{\rm FB}$, zero-crossing $q_0^2[A_{\rm FB}]$)
 \item $B^{(+)}\to K^{*(+)} \ell^+ \ell^-$: BR and angular observables ($A_{\rm FB}$, zero-crossing $q_0^2[A_{\rm FB}], F_L, A_T^i, P_i^{(\prime)}, S_i, R_{K^*}, R_{K^{*+}}$,...)
 \item $B_s\to \phi \ell^+ \ell^-$: BR and angular observables ($F_L, S_i, R_{\phi}$)
 \item $B^{(+)}\to K^{(+)} \ell^+ \ell^-$: BR and angular observables $A_{\rm FB}, F_H, R_{K}, R_{K_S^0}$ 
 \item $\Lambda_b^+ \to \Lambda \ell^+ \ell^-$: BR and angular observables $A_{\rm FB}, F_L$ 
 \item $K^+ \to \pi^+ \nu \bar\nu$: BR
 \item $K_L \to \pi^0 \nu \bar\nu$: BR
\end{itemize}
Furthermore, BR($K_L\to \pi^0 \ell^+ \ell^-$) as well as the
$\Delta S = 2$ and $\Delta B = 2$ meson-mixing observables are included in the development version of \texttt{SuperIso}.

\section{Leptonic and semi-leptonic kaon decays}\label{sec:kaon}
As an example of the observables included in \texttt{SuperIso}, we describe below with more detail the leptonic and semi-leptonic kaon decays which have been implemented more recently.
To describe these decays, we employ an effective Hamiltonian similar to the one used for $B$-physics:

\begin{equation}
\mathcal{H}_{\rm eff}=-\frac{4G_F}{\sqrt{2}}\lambda_t^{sd}\frac{\alpha_e}{4\pi}\sum_k \left( C_k^{sd,\ell}O_k^{sd,\ell} + C_{Q_k}^{sd,\ell}Q_k^{sd,\ell}\right)\,,
\end{equation}
with $\lambda_t^{sd}\equiv V_{td}V_{ts}^*$ where the four-fermion operators are given by
\begin{align}
&O_9^{sd,\ell(\prime)} = (\bar{s} \gamma_\mu P_{L(R)} d)\,(\bar{\ell}\gamma^\mu \ell),
&&O_{10}^{sd,\ell(\prime)} = (\bar{s} \gamma_\mu P_{L(R)} d)\,(\bar{\ell}\gamma^\mu\gamma_5 \ell),
\\
&Q_{1}^{sd,\ell(\prime)} = (\bar{s} P_{R(L)} d)\,(\bar{\ell} \ell),
&&Q_{2}^{sd,\ell(\prime)} = (\bar{s} P_{R(L)} d)\,(\bar{\ell} \gamma_5 \ell),
\end{align}
\begin{equation}
  O_{L(R)}^{sd,\ell} = (\bar{s} \gamma_\mu P_{L(R)} d)\,(\bar{\nu}\gamma^\mu(1-\gamma_5) \nu),
\end{equation}
with $\ell$ referring to the lepton flavour. 
The corresponding Wilson coefficients $C_k^{sd,\ell(\prime)}$ contain the short-distance effects from the SM as well as from new physics
\begin{equation}
C_k^{sd,\ell} = C_{k,{\rm SM}}^{sd,\ell}+ C_{k,{\rm NP}}^{sd,\ell}.
\end{equation}

\subsection{\texorpdfstring{Branching ratio of $K \to \pi \nu \bar\nu$}{Branching ratio of K -> pi nu nu}}

The branching fractions of the $K^+\to \pi^+ \nu \bar{\nu}$ and $K_L\to \pi^0 \nu \bar{\nu}$ decays  with a sum over all neutrino flavours are given by~\cite{Bobeth:2016llm,Bobeth:2017ecx}
\begin{align}
  \label{eq:Br-Kppipnunu}
  {\rm BR}(K^+ \to \pi^+ \nu \bar{\nu}) & =
  \frac{\kappa_+ (1 + \Delta_{\rm EM})}{\lambda^{10}}\frac{1}{3} s_W^4 \sum_{\nu_\ell}
  \left[  {\rm Im}^2 \Big(\lambda_t C_L^{\nu_\ell} \Big)
        + {\rm Re}^2 \Big(-\lambda^{sd}_c\frac{ \lambda^4 P_c(X)}{s_W^2}
                         + \lambda^{sd}_t C_L^{\nu_\ell} \Big)\right] ,\\
  \label{eq:Br-KLpinunu}
  {\rm BR}(K_L \to \pi^0 \nu \bar{\nu}) & =  \frac{\kappa_L }{\lambda^{10}}\frac{1}{3}s_W^4 \sum_{\nu_\ell}
  {\rm Im}^2 \left[\lambda_t C_L^{\nu_\ell} \right] ,
\end{align}
and in order to include  the chirality flipped contributions, the Wilson coefficients should be replaced by $C_i \to C_i + C_i^\prime$.
In eq.~\ref{eq:Br-Kppipnunu} above, $P_c(X) = P_c^{\rm SD}(X) + \delta P_{c,u}$ where $P_c^{\rm SD}(X)$ corresponds to the short-distance charm contributions calculated at NNLO accuracy~\cite{Brod:2008ss} and $\delta P_{c,u}$ refers to long-distance contributions~\cite{Isidori:2003ts}.
The electromagnetic radiative correction due to photon exchanges is given by $\Delta_{\rm EM}=-0.003$ 
for $E^\gamma_{\rm max}\approx20$ MeV, and the factors $\kappa_{+,L}$ are~\cite{Mescia:2007kn}
\begin{align}\label{kappa}
\kappa_L &= (2.231\pm 0.013 )\cdot 10^{-10}\left[\frac{\lambda}{0.225}\right]^8, \\
\kappa_+ &= (0.5173\pm 0.0025 )\cdot 10^{-10}\left[\frac{\lambda}{0.225}\right]^8.
\end{align}

\begin{figure}
\begin{center}
\includegraphics[width=0.8\textwidth]{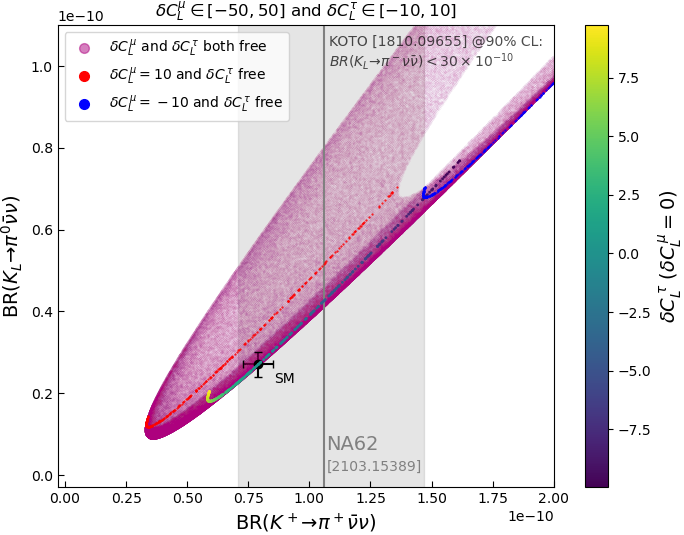}
\caption{
BR($K^+\to \pi^+ \bar{\nu}\nu$) vs BR($K_L\to \pi^0 \bar{\nu}\nu$) varying  $\delta C_L^\mu \in [-10,10]$ and $\delta C_L^\tau \in [-50,50]$. 
\label{fig:Kpinunu_CLmu10CLtau50}}
\end{center}
\end{figure}

The $K \to \pi \nu \bar\nu$ branching ratios are powerful probes of short-distance effects in the kaon sector as their SM predictions are very precise with $\lesssim$10\% uncertainty.
The effect of generic lepton flavour violating new physics contributions in the muon and tau sectors  can be seen in Fig.~\ref{fig:Kpinunu_CLmu10CLtau50} where $\delta C_L^\mu \in [-10,10]$ and $\delta C_L^\tau \in [-50,50]$. On the experimental side, for BR($K_L\to \pi^0 \nu \bar{\nu}$) there is currently only an upper bound from KOTO~\cite{KOTO:2018dsc} which is two orders of magnitude larger than the SM prediction. The main constraint is from BR($K^+\to \pi^+ \nu \bar{\nu}$) measured by NA62 collaboration~\cite{NA62:2021zjw} in agreement with the SM prediction which is $(7.86 \pm 0.61)\times 10^{-11}$.  
The red (blue) curve corresponds to the case where $\delta C_L^\tau$ is varied while $\delta C_L^\mu$ is fixed to $10\; (-10)$. The effect on the two branching ratios when having NP  effects solely in the tau sector can be seen with the color map (putting~$\delta C_L^\mu=0$).
%%%%%%%%%%%
%%%%%%%%%%%

\subsection{\texorpdfstring{Branching ratio of $K_{L,S} \to \mu \mu$}{Branching ratio of K(L,S) -> mu mu}}
The branching fractions of the $K_{S} \to \mu^+ \mu^-$ and $K_{L} \to \mu^+ \mu^-$ decays, are given by~\cite{Isidori:2003ts,Chobanova:2017rkj} 
\begin{align}
 \label{eq:brKSmumuComplete}
&{\rm BR}(K^0_{S} \to \mu^+ \mu^- ) =  \tau_{S}  \frac{ f_K^2 m_K^3 \beta_{\mu,K}} { 16 \pi} \left( \frac{G_F \alpha_e}{\sqrt{2}\pi}\right)^2 
\times \Bigg\{ \beta_{\mu,K}^2\left| \frac{\sqrt{2} \pi}{G_F \alpha_e} N_{S}^{\rm LD} - \frac{ m_K}{m_s +m_d}\mbox{Re}\big(\lambda_t C_{Q_1}\big)  \right|^2  \nonumber\\[-4pt]
&\qquad\qquad+\Bigg| \frac{2m_\mu}{m_K} \mbox{Im}\!\left(-\lambda_c\frac{ Y_c}{s_W^2} + \lambda_t C_{10}\right) + \frac{ m_K}{m_s +m_d}\mbox{Im}\big(\lambda_t C_{Q_2}\big)  \Bigg|^2
 \Bigg\},\\[4pt]
 \label{eq:brKLmumuComplete}
&{\rm BR}(K^0_{L} \to \mu^+ \mu^- ) =  \tau_{L}  \frac{ f_K^2 m_K^3 \beta_{\mu,K}} { 16 \pi} \left( \frac{G_F \alpha_e}{\sqrt{2}\pi}\right)^2 
\times \Bigg\{ \beta_{\mu,K}^2\left|  \frac{ m_K}{m_s +m_d}\mbox{Im}\big(\lambda_t C_{Q_1}\big)  \right|^2  \nonumber\\[-4pt]
&\qquad\qquad+\Bigg| \frac{\sqrt{2} \pi}{G_F \alpha_e} N_{L}^{\rm LD} - \frac{2m_\mu}{m_K} \mbox{Re}\!\left(-\lambda_c\frac{ Y_c}{s_W^2} + \lambda_t C_{10}\right) - \frac{ m_K}{m_s +m_d}\mbox{Re}\big(\lambda_t C_{Q_2}\big)  \Bigg|^2
 \Bigg\},
\end{align}
with $\beta_{\mu,K} \equiv \sqrt{1-4m_\mu/M_K^2}$.
In order to include  chirality flipped contributions, the Wilson coefficients should be replaced by $C_i \to C_i - C_i^\prime$.
The short-distance SM contributions are given by 
$C_{10}^{\rm SM}$
and the charm contribution $Y_c \left(\equiv \lambda^4 P_c(Y) \right)$, known with NNLO accuracy~\cite{Gorbahn:2006bm}. The long-distance contributions are denoted as $N_{S,L}^{\rm LD}$~\cite{Chobanova:2017rkj}
\begin{align}
N_{S}^{\rm LD} & =  (-2.65 + 1.14 i )\times 10^{-11} \textrm{\,(GeV})^{-2},\\
N_{L}^{\rm LD} & = \pm   \left[0.54(77) - 3.95 i\right]\times 10^{-11} \textrm{\,(GeV})^{-2},
\label{eq:LDKmumu}
\end{align}
where in the latter equation the sign is unclear since theoretically as well as experimentally the sign of the intermediate $2\gamma$ contribution due to ${\cal A}(K_L \to (\pi^0)^* \to \gamma \gamma)$ remains unknown. 
%%%%%%%%%%%
\begin{figure}[th!]
\begin{center}
\includegraphics[width=0.47\textwidth]{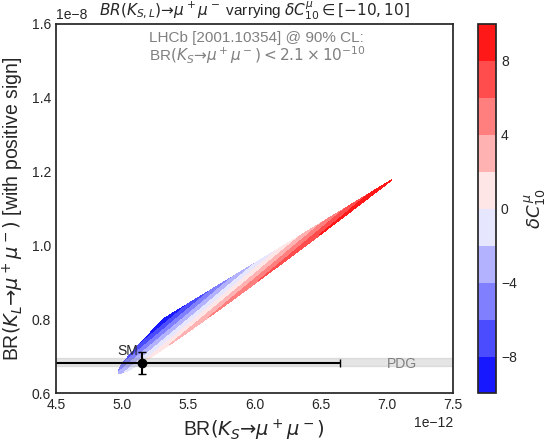}\qquad
\includegraphics[width=0.47\textwidth]{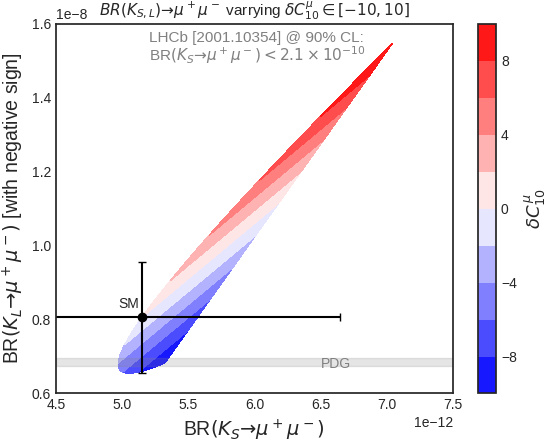}
% \vspace{-0.3cm}
\caption{
BR($K_S \to \bar{\mu}\mu$) vs BR($K_L\to \bar{\mu}\mu$) with positive (negative) sign for the long-distance contribution to $K_L\to \bar{\mu}\mu$ on the left (right) panel. 
\label{fig:Kmumu}}
\end{center}
\end{figure}
%%%%%%%%%%%

The effect of NP contributions from $\delta C_{10}\in[-10, 10]$ on BR($K_{L,S} \to \mu^+ \mu^-$) can be seen in Fig.~\ref{fig:Kmumu}.
On the experimental side, BR($K_L \to \mu^+ \mu^-$) is very precisely known~\cite{ParticleDataGroup:2020ssz} with an uncertainty of less than 2\%, however, considering the theoretical uncertainty of long-distance contributions and the sign ambiguity, there is room for NP effects.
On the other hand, for the $K_{S} \to \mu^+ \mu^-$ which is a much rarer decay, with three orders of magnitude suppression compared to the $K_L \to \mu^+ \mu^-$ in the SM, there is currently only an upper bound from LHCb~\cite{LHCb:2020ycd} which is about two orders of magnitude larger than the SM prediction.

For a detailed analysis of NP phenomenology with rare kaon decays, especially for scenarios involving lepton flavour universality violation (LFUV) see Ref.~\cite{DAmbrosio:2022kvb} where in addition to the above decay modes, the $K_L \to \pi^0 \ell^+ \ell^-$ decays as well as LFUV via $K^+ \to \pi^+ \ell^+ \ell^-$ have been considered. Furthermore, in Ref.~\cite{DAmbrosio:2022kvb} the overall NP fit to the rare kaon decays has been presented.

\section{\texorpdfstring{Global fit to rare $B$-decays}{Global fit to rare B-decays}}\label{sec:Bdecays}

%%%%%%%%%%%
\begin{figure}[b!]
\begin{center}
\includegraphics[width=0.55\textwidth]{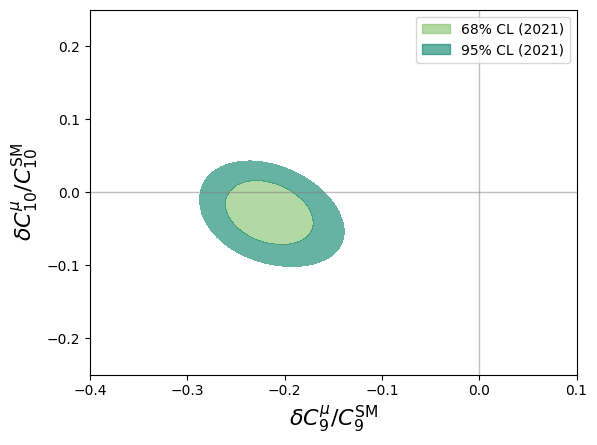}\\
\includegraphics[width=0.47\textwidth]{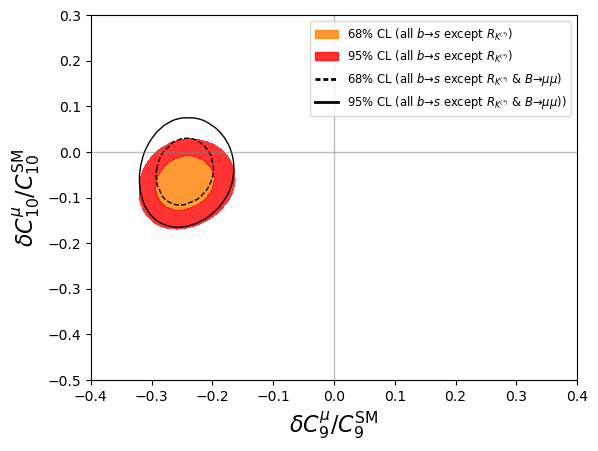}\qquad
\includegraphics[width=0.47\textwidth]{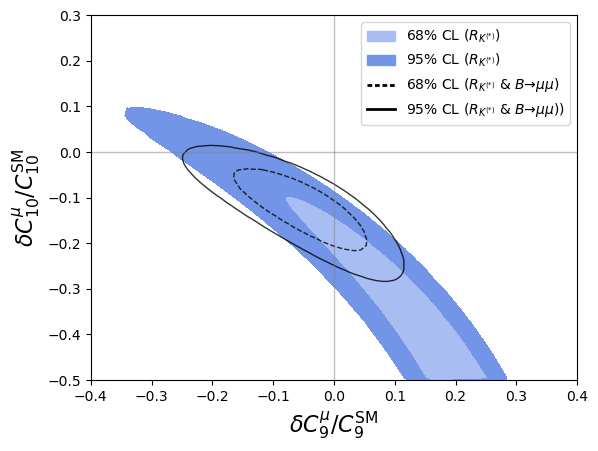}
% \vspace{-0.3cm}
\caption{
Two operator fits to $(C_9^\mu, C_{10}^\mu)$ within 68 and 95\% confidence level. 
The upper plot corresponds to  the fit to all observables (assuming 10\% power corrections).
In the lower plots, on  the left  we have considered all observables except $R_K$ and $R_{K^*}$ (with the assumption of 10\% power corrections) while on the right we have only used the data on $R_K, R_{K^*}$. Further details can be found in Ref.~\cite{Hurth:2021nsi}.
\label{fig:bsllfit}}
\end{center}
\end{figure}
%%%%%%%%%%%

In recent years, the most promising signs of new physics have been in rare $B$-decays where several observables have been measured to be in tension with their corresponding SM predictions. 
However, there are numerous $B$-physics observables, inter-dependent via one or more Wilson coefficients and in order to get a consistent global picture of the implications $B$-physics measurements, it is suitable to perform a global fit to all observables which are described via common Wilson coefficients. This can be done by doing a $\chi^2$ fit to the relevant Wilson coefficients of the $B$-physics observables.
In \texttt{SuperIso}, there are dedicated subroutines in order to calculate 
\begin{align}
 \chi^2 =  \sum_{i,j =1}^{N} \left( O_i^{\rm th} - O_i^{\rm exp} \right)
 \; C_{ij}^{-1} \; \left( O_j^{\rm th} - O_j^{\rm exp} \right)\,,
\end{align}
where $O^{\rm exp}_i$ and $O^{\rm th}_i$ denote the experimental measurement and theoretical predictions of the $i^{\rm th}$ observable, respectively.   
$C_{i,j} $ describes the covariance matrix which is the sum of the theoretical and experimental covariance matrices. In \texttt{SuperIso} $C_{i,j} $ is calculated automatically for the SM and then used for all NP points. It is also possible to calculate $C_{i,j} $ for each NP point, however, computationally this would be  very demanding and as shown in Ref.~\cite{Bhom:2020lmk}, the assumption that $C_{i,j} $ calculated for the SM can be used also for new physics points is a justified approximation.

In Fig.~\ref{fig:bsllfit}, two-dimensional fits to Wilson coefficients $(C_{9}^\mu, C_{10}^\mu)$ are shown where in the first row all the relevant $b\to s$ data have been used while in the second row separate fits to the lepton flavour universality violating/conserving observables have been considered, showing the coherence of the two sets of data with the global fit.

\section{Conclusions}\label{sec:conclusion}
Flavour observables offer powerful probes of physics beyond the Standard Model and can explore energy scales not reachable via direct detection.
\texttt{SuperIso} calculates most of the constraining flavour observables
and it also includes many other observables such as muon  anomalous magnetic
moment and the dark matter constraints.
Here we presented only a few examples in model-independent new physics scenarios,
but \texttt{SuperIso} also automatically calculates the Wilson coefficients of several model-specific scenarios such as two Higgs doublet models (2HDM), minimal supersymmetric Standard
Model (MSSM) and next-to-minimal supersymmetric Standard Model (NMSSM). 
In the next versions, \texttt{SuperIso}  is going to be extended to include further flavour observables as well as the automatic calculation of Wilson coefficients of non-minimal flavour violating MSSM scenarios~\cite{Boussejra:2022heb}. Furthermore, there is also going to be a direct interface
to \texttt{MARTY}~\cite{Uhlrich:2020ltd} which is a C++ public program that calculates Wilson coefficients for generic
BSM scenarios at one-loop level.

\acknowledgments
The speaker, SN, would like to thank the organisers of the CompTools2021 workshop for their kind invitation and for providing the opportunity to present this~work.  SN was supported in part by the INFN research initiative Exploring New Physics~(ENP).

\providecommand{\href}[2]{#2}\begingroup\raggedright\endgroup


\begin{thebibliography}{10}

\bibitem{Mahmoudi:2007vz}
F.~Mahmoudi, \emph{{SuperIso: A Program for calculating the isospin asymmetry
  of $B \to K^* \gamma$ in the MSSM}},
  \href{https://doi.org/10.1016/j.cpc.2007.12.006}{\emph{Comput. Phys. Commun.}
  {\bfseries 178} (2008) 745}
  [\href{https://arxiv.org/abs/0710.2067}{{\ttfamily 0710.2067}}].

\bibitem{Mahmoudi:2008tp}
F.~Mahmoudi, \emph{{SuperIso v2.3: A Program for calculating flavor physics
  observables in Supersymmetry}},
  \href{https://doi.org/10.1016/j.cpc.2009.02.017}{\emph{Comput. Phys. Commun.}
  {\bfseries 180} (2009) 1579}
  [\href{https://arxiv.org/abs/0808.3144}{{\ttfamily 0808.3144}}].

\bibitem{Mahmoudi:2009zz}
F.~Mahmoudi, \emph{{SuperIso v3.0, flavor physics observables calculations:
  Extension to NMSSM}},
  \href{https://doi.org/10.1016/j.cpc.2009.05.001}{\emph{Comput. Phys. Commun.}
  {\bfseries 180} (2009) 1718}.

\bibitem{Neshatpour:2021nbn}
S.~Neshatpour and F.~Mahmoudi, \emph{{Flavour Physics with SuperIso}},
  \href{https://doi.org/10.22323/1.392.0036}{\emph{PoS} {\bfseries TOOLS2020}
  (2021) 036} [\href{https://arxiv.org/abs/2105.03428}{{\ttfamily
  2105.03428}}].

\bibitem{Arbey:2009gu}
A.~Arbey and F.~Mahmoudi, \emph{{SuperIso Relic: A Program for calculating
  relic density and flavor physics observables in Supersymmetry}},
  \href{https://doi.org/10.1016/j.cpc.2010.03.010}{\emph{Comput. Phys. Commun.}
  {\bfseries 181} (2010) 1277}
  [\href{https://arxiv.org/abs/0906.0369}{{\ttfamily 0906.0369}}].

\bibitem{Arbey:2011zz}
A.~Arbey and F.~Mahmoudi, \emph{{SuperIso Relic v3.0: A program for calculating
  relic density and flavour physics observables: Extension to NMSSM}},
  \href{https://doi.org/10.1016/j.cpc.2011.03.019}{\emph{Comput. Phys. Commun.}
  {\bfseries 182} (2011) 1582}.

\bibitem{Arbey:2018msw}
A.~Arbey, F.~Mahmoudi and G.~Robbins, \emph{{SuperIso Relic v4: A program for
  calculating dark matter and flavour physics observables in Supersymmetry}},
  \href{https://doi.org/10.1016/j.cpc.2019.01.014}{\emph{Comput. Phys. Commun.}
  {\bfseries 239} (2019) 238}
  [\href{https://arxiv.org/abs/1806.11489}{{\ttfamily 1806.11489}}].

\bibitem{Bobeth:2016llm}
C.~Bobeth, A.J.~Buras, A.~Celis and M.~Jung, \emph{{Patterns of Flavour
  Violation in Models with Vector-Like Quarks}},
  \href{https://doi.org/10.1007/JHEP04(2017)079}{\emph{JHEP} {\bfseries 04}
  (2017) 079} [\href{https://arxiv.org/abs/1609.04783}{{\ttfamily
  1609.04783}}].

\bibitem{Bobeth:2017ecx}
C.~Bobeth and A.J.~Buras, \emph{{Leptoquarks meet $\varepsilon'/\varepsilon$
  and rare Kaon processes}},
  \href{https://doi.org/10.1007/JHEP02(2018)101}{\emph{JHEP} {\bfseries 02}
  (2018) 101} [\href{https://arxiv.org/abs/1712.01295}{{\ttfamily
  1712.01295}}].

\bibitem{Brod:2008ss}
J.~Brod and M.~Gorbahn, \emph{{Electroweak Corrections to the Charm Quark
  Contribution to $K^+ \to \pi^+ \nu \bar\nu$}},
  \href{https://doi.org/10.1103/PhysRevD.78.034006}{\emph{Phys. Rev. D}
  {\bfseries 78} (2008) 034006}
  [\href{https://arxiv.org/abs/0805.4119}{{\ttfamily 0805.4119}}].

\bibitem{Isidori:2003ts}
G.~Isidori and R.~Unterdorfer, \emph{{On the short distance constraints from
  $K(L,S) \to \mu^+ \mu^-$}},
  \href{https://doi.org/10.1088/1126-6708/2004/01/009}{\emph{JHEP} {\bfseries
  01} (2004) 009} [\href{https://arxiv.org/abs/hep-ph/0311084}{{\ttfamily
  hep-ph/0311084}}].

\bibitem{Mescia:2007kn}
F.~Mescia and C.~Smith, \emph{{Improved estimates of rare K decay
  matrix-elements from Kl3 decays}},
  \href{https://doi.org/10.1103/PhysRevD.76.034017}{\emph{Phys. Rev. D}
  {\bfseries 76} (2007) 034017}
  [\href{https://arxiv.org/abs/0705.2025}{{\ttfamily 0705.2025}}].

\bibitem{KOTO:2018dsc}
{\scshape KOTO} collaboration, \emph{{Search for the $K_L \!\to\! \pi^0 \nu
  \overline{\nu}$ and $K_L \!\to\! \pi^0 X^0$ decays at the J-PARC KOTO
  experiment}},
  \href{https://doi.org/10.1103/PhysRevLett.122.021802}{\emph{Phys. Rev. Lett.}
  {\bfseries 122} (2019) 021802}
  [\href{https://arxiv.org/abs/1810.09655}{{\ttfamily 1810.09655}}].

\bibitem{NA62:2021zjw}
{\scshape NA62} collaboration, \emph{{Measurement of the very rare
  K$^{+}$\textrightarrow{}$ {\pi}^{+}\nu \overline{\nu} $ decay}},
  \href{https://doi.org/10.1007/JHEP06(2021)093}{\emph{JHEP} {\bfseries 06}
  (2021) 093} [\href{https://arxiv.org/abs/2103.15389}{{\ttfamily
  2103.15389}}].

\bibitem{Chobanova:2017rkj}
V.~Chobanova, G.~D'Ambrosio, T.~Kitahara, M.~Lucio~Martinez,
  D.~Martinez~Santos, I.S.~Fernandez et~al., \emph{{Probing SUSY effects in
  $K_S^0\rightarrow\mu^+\mu^-$}},
  \href{https://doi.org/10.1007/JHEP05(2018)024}{\emph{JHEP} {\bfseries 05}
  (2018) 024} [\href{https://arxiv.org/abs/1711.11030}{{\ttfamily
  1711.11030}}].

\bibitem{Gorbahn:2006bm}
M.~Gorbahn and U.~Haisch, \emph{{Charm Quark Contribution to $K(L) \to \mu^+
  \mu^-$ at Next-to-Next-to-Leading}},
  \href{https://doi.org/10.1103/PhysRevLett.97.122002}{\emph{Phys. Rev. Lett.}
  {\bfseries 97} (2006) 122002}
  [\href{https://arxiv.org/abs/hep-ph/0605203}{{\ttfamily hep-ph/0605203}}].

\bibitem{ParticleDataGroup:2020ssz}
{\scshape Particle Data Group} collaboration, \emph{{Review of Particle
  Physics}}, \href{https://doi.org/10.1093/ptep/ptaa104}{\emph{PTEP} {\bfseries
  2020} (2020) 083C01}.

\bibitem{LHCb:2020ycd}
{\scshape LHCb} collaboration, \emph{{Constraints on the $K^0_S \rightarrow
  \mu^+ \mu^-$ Branching Fraction}},
  \href{https://doi.org/10.1103/PhysRevLett.125.231801}{\emph{Phys. Rev. Lett.}
  {\bfseries 125} (2020) 231801}
  [\href{https://arxiv.org/abs/2001.10354}{{\ttfamily 2001.10354}}].

\bibitem{DAmbrosio:2022kvb}
G.~D'Ambrosio, A.M.~Iyer, F.~Mahmoudi and S.~Neshatpour, \emph{{Anatomy of kaon
  decays and prospects for lepton flavour universality violation}},
  \href{https://arxiv.org/abs/2206.14748}{{\ttfamily 2206.14748}}.

\bibitem{Hurth:2021nsi}
T.~Hurth, F.~Mahmoudi, D.M.~Santos and S.~Neshatpour, \emph{More indications
  for lepton nonuniversality in $b \to s \ell^+\ell^-$},
  \href{https://doi.org/10.1016/j.physletb.2021.136838}{\emph{Phys. Lett. B}
  {\bfseries 824} (2022) 136838}
  [\href{https://arxiv.org/abs/2104.10058}{{\ttfamily 2104.10058}}].

\bibitem{Bhom:2020lmk}
J.~Bhom, M.~Chrzaszcz, F.~Mahmoudi, M.T.~Prim, P.~Scott and M.~White, \emph{{A
  model-independent analysis of $b{\rightarrow }s\mu ^{+}\mu ^{-}$transitions
  with GAMBIT \textquoteright{}s FlavBit}},
  \href{https://doi.org/10.1140/epjc/s10052-021-09840-z}{\emph{Eur. Phys. J. C}
  {\bfseries 81} (2021) 1076}
  [\href{https://arxiv.org/abs/2006.03489}{{\ttfamily 2006.03489}}].

\bibitem{Boussejra:2022heb}
M.A.~Boussejra, F.~Mahmoudi and G.~Uhlrich, \emph{{Flavour anomalies in
  supersymmetric scenarios with non-minimal flavour violation}},
  \href{https://arxiv.org/abs/2201.04659}{{\ttfamily 2201.04659}}.

\bibitem{Uhlrich:2020ltd}
G.~Uhlrich, F.~Mahmoudi and A.~Arbey, \emph{{MARTY - Modern ARtificial
  Theoretical phYsicist A C++ framework automating symbolic calculations Beyond
  the Standard Model}},
  \href{https://doi.org/10.1016/j.cpc.2021.107928}{\emph{Comput. Phys. Commun.}
  {\bfseries 264} (2021) 107928}
  [\href{https://arxiv.org/abs/2011.02478}{{\ttfamily 2011.02478}}].

\end{thebibliography}
\end{document}